\documentclass[aps,prl,twocolumn,superscriptaddress,showpacs]{revtex4}
\def\mystyle{2}

\if\mystyle4
 \def\figonescale{0.7}
 \def\figtwoscale{0.8}
 
  \def\bellelogo{\vbox to 16mm{
                 \vss\hbox{\resizebox{!}{3cm}{
                 \includegraphics{belle-logo.eps}}}}\vspace{-1cm}}
  \def\preprintA{\hbox{\hfil BELLE-CONF-0401}}
  \def\preprintB{\hbox{\hfil ICHEP 11-0646}}
  \def\preprintC{}
\else
 \def\figonescale{0.4}
 \def\figtwoscale{0.48}
 
 \def\bellelogo{}
 \def\preprintA{}
 \def\preprintB{}
 \def\preprintC{}
\fi

\def\mydate{\date{May 30 2005}}

\usepackage{graphicx}

\def\KM{K^-}

\def\KS{{K^0_S}}

\def\piZ{{\pi^0}}
\def\piP{{\pi^+}}
\def\piM{{\pi^-}}

\def\rhoM{{\rho^-}}
\def\rhoZ{{\rho^0}}
\def\Kstar{{K^*}}

\def\KstarM{{K^{*-}}}
\def\KstarB{{\Kbar^{*0}}}
\def\omegaG{{\omega\gamma}}

\def\rhoMG{{\rhoM\gamma}}
\def\rhoZG{{\rhoZ\gamma}}
\def\ROG{{(\rho,\omega)\gamma}}
\def\KstarG{{\Kstar\gamma}}

\def\KstarMG{{\KstarM\gamma}}

\def\qqbar{q\overline{q}{}}

\def\Bbar{\overline{B}{}}

\def\Kbar{\overline{K}{}}

\def\epem{e^+e^-{}}

\def\btosgamma{b\to s\gamma}

\def\btodgamma{b\to d\gamma}

\def\BtoRG{B\to \rho\gamma}

\def\BtoRMG{B^-\to \rho^-\gamma}
\def\BtoRZG{B^0\to \rho^0\gamma}
\def\BtoRBG{\Bbar^0\to \rho^0\gamma}
\def\BtoROG{B\to (\rho,\omega)\gamma}
\def\BtoOG{\Bbar^0\to \omega\gamma}

\def\BtoKG{B\to K^*\gamma}

\def\BtoXsgamma{B\to X_s\gamma}

\def\BtoKBG{\Bbar^0\to \Kbar^{*0}\gamma}
\def\BtoKMG{B^-\to K^{*-}\gamma}

\def\BtoDZpi{B^-\to D^0\pi^-}

\def\cm{\mbox{~cm}}

\def\GeV{\mbox{~GeV}}
\def\GeVc{\mbox{~GeV}/c}
\def\GeVcc{\mbox{~GeV}/c^2}

\def\MeVc{\mbox{~MeV}/c}
\def\MeVcc{\mbox{~MeV}/c^2}

\def\Vtd{V_{td}}
\def\Vts{V_{ts}}

\def\Br{{\cal B}}

\def\Lpi{L_\pi}
\def\LK{L_K}

\def\Mbc{M_{\rm bc}}
\def\DeltaE{\Delta{E}}
\def\Ebeam{E^*_{\rm beam}{}}

\def\MKpi{M_{K\pi}}

\def\Egamma{E^*_\gamma}

\def\tauBratio{{\tau_{B^+}\over\tau_{B^0}}}

\def\piZeta{\piZ/\eta}

\def\thetaB{{\theta^*_B}}
\def\cosB{{\cos\thetaB}}

\def\calF{{\cal F}}
\def\calLs{{\cal L}_s}
\def\calLc{{\cal L}_c}
\def\calR{{\cal R}}
\def\calL{{\cal L}}

\def\Deltaz{\Delta{z}}
\def\thetahel{\theta_{\rm hel}}
\def\coshel{\cos\thetahel}

\def\Lzero{{\cal L}_0}
\def\Lmax{{\cal L}_{\rm max}}

\def\PM#1#2{\,^{+#1}_{-#2}{}}
\def\EM#1{\times10^{-#1}}

\def\etal{\textit{et al.}}
\def\Journal#1#2#3#4{{#1} {\bf #2}, #3 (#4)} 
\def\NIMA{Nucl. Instrum. Meth. A}
\def\NPB{Nucl. Phys. B}
\def\PLB{Phys. Lett. B}
\def\PRL{Phys. Rev. Lett.}
\def\PRD{Phys. Rev. D}
\def\ZPC{Z. Phys. C}

\def\EPJC{Eur. Phys. J. C}
\def\JPG{J. Phys. G}

\makeatletter
\def\mygraphic@[#1,#2]#3{\resizebox{#1}{#2}{\includegraphics{#3}}}
\def\mygraphics{\@ifnextchar[{\mygraphic@}{\mygraphic@[\textwidth,!]}}
\def\myep@[#1]#2{\resizebox{#1\textwidth}{!}{\includegraphics{#2}}}
\def\myeps{\@ifnextchar[{\myep@}{\myep@[1]}}
\makeatother

\def\EffKaon{76\mbox{--}80\%} 
\def\EffPionR{89\%}  

\def\EffPionO{94\%}  
\def\FakePionR{{\sim}10\%} 

\def\EffVtxVeryRough{85\%} 

\def\drcut{2\cm}
\def\dzcut{5\cm}
\def\ptrkcut{100\MeVc}

\def\effRPG{(5.5\pm0.4)\%}
\def\effRZG{(3.9\pm0.3)\%}
\def\effOMG{(3.9\pm0.4)\%}

\def\sROG{1.2}
\def\sRPG{1.7}
\def\sRZG{0.3}
\def\sOMG{0.1}

\def\BrROG{(0.72\PM{0.43}{0.39}\pm{0.28})\EM6}
\def\BrRPG{(1.24\PM{0.67}{0.61}\pm{0.26})\EM6}
\def\BrRZG{(0.18\PM{0.33}{0.25}\pm{0.10})\EM6}
\def\BrOMG{(0.08\PM{0.39}{0.31}\pm{0.19})\EM6}

\def\nRPG{18.7\PM{10.1}{\phantom{1}9.2}}
\def\nRZG{\phantom{1} 1.9\PM{2.8}{2.7\phantom{1}}}
\def\nOMG{\phantom{1} 0.9\PM{4.2}{3.3\phantom{1}}}

\def\ULROG{1.4\EM6}
\def\ULRPG{2.2\EM6}
\def\ULRZG{0.8\EM6}
\def\ULOMG{0.8\EM6}

\def\ULROGoverKG{0.035}

\def\ULVtdoVts{0.22}

\begin{document}


\bellelogo

\preprint{\vbox{
  \preprintA
  \preprintB
  \preprintC
}}

\title{Search for the \boldmath$\btodgamma$ process}


\affiliation{Budker Institute of Nuclear Physics, Novosibirsk}
\affiliation{Chiba University, Chiba}
\affiliation{Chonnam National University, Kwangju}
\affiliation{University of Cincinnati, Cincinnati, Ohio 45221}
\affiliation{Gyeongsang National University, Chinju}
\affiliation{University of Hawaii, Honolulu, Hawaii 96822}
\affiliation{High Energy Accelerator Research Organization (KEK), Tsukuba}
\affiliation{Hiroshima Institute of Technology, Hiroshima}
\affiliation{Institute of High Energy Physics, Chinese Academy of Sciences, Beijing}
\affiliation{Institute of High Energy Physics, Vienna}
\affiliation{Institute for Theoretical and Experimental Physics, Moscow}
\affiliation{J. Stefan Institute, Ljubljana}
\affiliation{Kanagawa University, Yokohama}
\affiliation{Korea University, Seoul}
\affiliation{Kyungpook National University, Taegu}
\affiliation{Swiss Federal Institute of Technology of Lausanne, EPFL, Lausanne}
\affiliation{University of Ljubljana, Ljubljana}
\affiliation{University of Melbourne, Victoria}
\affiliation{Nagoya University, Nagoya}
\affiliation{Nara Women's University, Nara}
\affiliation{National Central University, Chung-li}
\affiliation{National United University, Miao Li}
\affiliation{Department of Physics, National Taiwan University, Taipei}
\affiliation{H. Niewodniczanski Institute of Nuclear Physics, Krakow}
\affiliation{Nippon Dental University, Niigata}
\affiliation{Niigata University, Niigata}
\affiliation{Nova Gorica Polytechnic, Nova Gorica}
\affiliation{Osaka City University, Osaka}
\affiliation{Osaka University, Osaka}
\affiliation{Panjab University, Chandigarh}
\affiliation{Peking University, Beijing}
\affiliation{Princeton University, Princeton, New Jersey 08544}
\affiliation{University of Science and Technology of China, Hefei}
\affiliation{Seoul National University, Seoul}
\affiliation{Shinshu University, Nagano}
\affiliation{Sungkyunkwan University, Suwon}
\affiliation{University of Sydney, Sydney NSW}
\affiliation{Tata Institute of Fundamental Research, Bombay}
\affiliation{Toho University, Funabashi}
\affiliation{Tohoku Gakuin University, Tagajo}
\affiliation{Tohoku University, Sendai}
\affiliation{Department of Physics, University of Tokyo, Tokyo}
\affiliation{Tokyo Institute of Technology, Tokyo}
\affiliation{Tokyo Metropolitan University, Tokyo}
\affiliation{Tokyo University of Agriculture and Technology, Tokyo}
\affiliation{University of Tsukuba, Tsukuba}
\affiliation{Virginia Polytechnic Institute and State University, Blacksburg, Virginia 24061}
\affiliation{Yonsei University, Seoul}
   \author{D.~Mohapatra}\affiliation{Virginia Polytechnic Institute and State University, Blacksburg, Virginia 24061} 
   \author{M.~Nakao}\affiliation{High Energy Accelerator Research Organization (KEK), Tsukuba} 
   \author{B.~D.~Yabsley}\affiliation{Virginia Polytechnic Institute and State University, Blacksburg, Virginia 24061} 
 \author{K.~Abe}\affiliation{High Energy Accelerator Research Organization (KEK), Tsukuba} 
   \author{K.~Abe}\affiliation{Tohoku Gakuin University, Tagajo} 
   \author{H.~Aihara}\affiliation{Department of Physics, University of Tokyo, Tokyo} 
   \author{K.~Arinstein}\affiliation{Budker Institute of Nuclear Physics, Novosibirsk} 
   \author{Y.~Asano}\affiliation{University of Tsukuba, Tsukuba} 
   \author{T.~Aushev}\affiliation{Institute for Theoretical and Experimental Physics, Moscow} 
   \author{T.~Aziz}\affiliation{Tata Institute of Fundamental Research, Bombay} 
% ...
\collaboration{The Belle Collaboration}


\mydate

\begin{abstract} 

  We report the results of a search for the flavor-changing neutral
  current process $\btodgamma$ using a sample of 275 million $B$ meson
  pairs accumulated by the Belle detector at KEKB.  We find no
  significant signal for the exclusive decays $\BtoRMG$, $\BtoRBG$, or
  $\BtoOG$. Assuming an isospin relation between the three modes, we
  set an upper limit for the combined branching fraction
  $\Br(\BtoROG)$, of $\ULROG$ at the 90\% confidence level.  This
  limit can be used to constrain the ratio of CKM matrix elements
  $|\Vtd/\Vts|$.

\end{abstract}



\maketitle



\begin{figure}[ht]
\begin{center}
\vspace*{24pt}
\myeps[\figonescale]{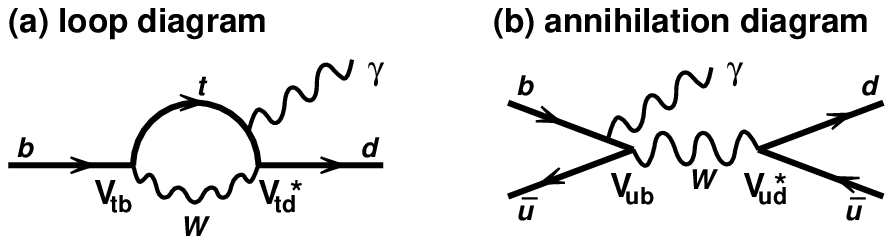}
\caption{(a) Loop diagram for $\btodgamma$ and (b) annihilation diagram
  for $\BtoRMG$ only.}
\label{fig:diagram}
\end{center}
\end{figure}

The $\btodgamma$ process, shown in Fig.~\ref{fig:diagram}(a), is a
flavor changing neutral current transition that proceeds via loop
diagrams in the Standard Model (SM).  It is suppressed with respect to
$\btosgamma$ by the Cabibbo-Kobayashi-Maskawa (CKM) factor
$|\Vtd/\Vts|^2 \sim 0.04$ with a large uncertainty due to the lack of
precise knowledge of $\Vtd$. The exclusive modes $B \to \rho \gamma$
and $B \to \omega \gamma$, which are presumably the easiest modes to
search for, have not yet been observed~\cite{bib:babar-rhogam}.
Calculations based on the measured rate for the $\btosgamma$ process
$\BtoKG$ that include $|\Vtd/\Vts|^2$ suppression, corrections due to
form factors, $SU(3)$ breaking effects, and, for $B^-$ decay, the
additional annihilation diagram of Fig.~\ref{fig:diagram}(b), predict
branching fractions in the range $(0.9\mbox{--}2.7)\EM6$
~\cite{bib:ali-parkhomenko,bib:bosch-buchalla}.  Measurement of these
exclusive branching fractions would improve the constraints on
$|\Vtd/\Vts|$ in the context of the SM, and provide sensitivity to
physics beyond the SM that is complementary to that from $\btosgamma$.


In this paper, we report the results of a search for the $\btodgamma$
process using a sample of $(275\pm3)$ million $B$ meson pairs
accumulated at the $\Upsilon(4S)$ resonance.  The data are produced in
$\epem$ annihilation at the KEKB energy-asymmetric (3.5 on 8 GeV)
collider~\cite{bib:kekb} and collected with the Belle
detector~\cite{bib:belle-detector}.  The Belle detector includes a
silicon vertex detector (SVD), a central drift chamber (CDC), aerogel
threshold Cherenkov counters (ACC), time-of-flight (TOF) scintillation
counters, and an electromagnetic calorimeter (ECL) comprised of CsI(Tl)
crystals located inside a 1.5 T superconducting solenoid coil.  An iron
flux-return located outside of the coil is instrumented to identify
muons (KLM).  The dataset consists of two subsets: the first 152 million
$B$ meson pairs were collected with a 2.0 cm radius beampipe and a 3-layer SVD, and the
remaining 123 million $B$ meson pairs with a 1.5 cm radius beampipe, a
4-layer SVD and a small-cell inner drift chamber~\cite{bib:svd2}.


We reconstruct the following final states: $\BtoRMG$, $\BtoRBG$, and
$\BtoOG$.  Charge conjugate modes are implicitly included throughout
this paper.  We also reconstruct control samples of $\BtoKMG$ and
$\BtoKBG$ decays.  The following decay chains are used to reconstruct
the intermediate states: $\rhoM\to\piM\piZ$, $\rhoZ\to\piP\piM$,
$\omega\to\piP\piM\piZ$, $\KstarM\to\KM\piZ$, $\KstarB\to\KM\piP$, and
$\piZ\to\gamma\gamma$.

Photon candidates are reconstructed from isolated clusters in the ECL
that have no corresponding charged track, and a shower shape that is
consistent with that of a photon.  A photon in the barrel region of
the ECL ($33^\circ<\theta_\gamma<128^\circ$ in the laboratory frame)
with a center-of-mass (c.m.)\ energy in the range
$1.8\GeV<\Egamma<3.4\GeV$ is selected as the primary photon candidate.
To suppress backgrounds from $\piZ\to\gamma\gamma$ and
$\eta\to\gamma\gamma$ decays, we veto the event if the reconstructed
mass of the primary photon and any other photon of energy 30 (200) MeV
or more is within a $\pm3\sigma$ window, i.e., $\pm18$ $(\pm
32)\MeVcc$, around the $\piZ$ ($\eta$) mass.  This set of criteria is
referred to as the $\piZeta$ veto.  For the primary photon, we sum the
energy deposited in arrays of $3\times3$ and $5\times5$ ECL cells
around the maximum energy cell; if their ratio is less than 0.95, the
event is vetoed.

Neutral pions are formed from photon pairs with invariant masses
within $\pm 10$ ($16$)$\MeVcc$ of the nominal $\piZ$ mass,
corresponding to a ${\sim}2 \sigma$ $({\sim}3\sigma)$ window for the
$\rhoMG$ and $\KstarMG$ ($\omegaG$) modes.  The photon momenta are
then recalculated with a $\piZ$ mass constraint.  We require the
energy of each photon to be greater than 30 MeV.  We also require the
{c.m.}  momentum of the $\piZ$ to be greater than $0.5\GeVc$ for the
$\rhoMG$ and $\KstarMG$ modes.

Charged pions and kaons are reconstructed as tracks in the CDC and
SVD.  Each track is required to have a transverse momentum greater
than $\ptrkcut$ and distance of closest approach to the interaction
point
of less than $\drcut$ in radius and $\pm\dzcut$ along the $z$-axis,
which is aligned opposite to the positron beam.  We do not use the
track to form the signal candidate if, when combined with any other
track, it forms a $\KS$ candidate with an invariant mass within
$\pm10\MeVcc$ of the nominal $\KS$ mass and a displaced vertex that is
consistent with a $\KS$.  We determine the pion ($\Lpi$) and kaon
($\LK$) likelihoods from ACC, CDC and TOF information, and form a
likelihood ratio $\Lpi/(\Lpi + \LK)$ to separate pions from kaons.  We
select pions using criteria which have an efficiency of $\EffPionR$
and for which $\FakePionR$ of kaons are misidentified as pions.  For
the $\omegaG$ mode, we relax the requirement to select $\EffPionO$ of
pions.  (In the $\KstarG$ modes, we select kaons with an efficiency of
$\EffKaon$.)  In addition, we remove tracks that are consistent with
being electrons or muons.

Invariant masses for the $\rho$ and $\omega$ candidates are required
to be within windows of $\pm150\MeVcc$ and $\pm30\MeVcc$,
respectively, around their nominal masses.

$B$ candidates are reconstructed by combining a $\rho$ or $\omega$
candidate and the primary photon using two variables: the beam-energy
constrained mass $\Mbc = \sqrt{ (\Ebeam/c^2)^2 - |p_{B}^*/c|^{2}}$ and
the energy difference $\Delta E = E^*_{B} - \Ebeam$; $p^*_{B}$ and
$E^*_B$ are the {c.m.}\ momentum and energy of the $B$ candidate, and
$\Ebeam$ is the {c.m.}\ beam energy.  The magnitude of the photon
momentum is replaced by $(\Ebeam - E_{\rho/\omega}^*)/c$ when the
momentum $p^*_{B}$ is calculated.  To optimize the event selection, we
study Monte Carlo (MC) events in the region
$-0.10\GeV<\DeltaE<0.08\GeV$ and $5.273\GeVcc<\Mbc<5.285\GeVcc$: we
choose selection criteria to maximize $N_S/\sqrt{N_B}$, where $N_S$ is
the MC signal yield assuming the SM branching fractions in
Ref.~\cite{bib:ali-parkhomenko}, and $N_B$ is the expected background
yield.


The dominant background arises from continuum events
($\epem\to\qqbar(\gamma)$), where the accidental combination of a
$\rho$ or $\omega$ candidate with a photon forms a $B$ candidate.  We
suppress this background using the following quantities:
(1) $\calF$, a Fisher discriminant 
\cite{bib:fisher} constructed from 16 modified Fox-Wolfram
moments~\cite{bib:belle-pi0pi0,bib:fox-wolfram} and the scalar sum of
the transverse momentum of all charged tracks and photons.
(2) $\cosB$, where $\thetaB$ is the {c.m.}\ polar angle of the $B$
candidate direction: true $B$ mesons follow a $1-\cos^2\thetaB$
distribution, while candidates in the continuum background are
uniformly distributed.
(3) $\Delta z$, the separation between the decay vertex of the candidate
$B$ meson and the origin of the remaining tracks in the event along the
$z$-axis.  The two vertices are 
reconstructed in about $\EffVtxVeryRough$ of
the $\rhoZG$ and $\omegaG$ events.  Discrimination is provided due to
the displacement of the signal $B$ decay vertex from the other $B$, 
as tracks from continuum events typically have a common vertex.
%
%
(4)
A tagging quality $r$ which indicates the level of confidence in the
$B$-flavor determination described in Ref.~\cite{bib:hamlet}.  The
algorithm provides additional discrimination between signal and
continuum background where no true $B$ meson is present.

For each of the quantities $\calF$, $\cosB$ and $\Deltaz$, we
construct one-dimensional likelihood distributions for signal and
continuum events.  Signal distributions are modeled with an asymmetric
Gaussian function for $\calF$, ${3\over2}(a_0-a_2\cos^2\thetaB)$ for
$\cos\thetaB$, and an exponential convolved with a Gaussian resolution
function for $\Deltaz$. Continuum background distributions are modeled
with an asymmetric Gaussian for $\calF$, $(b_0-b_2\cos^2\thetaB)$ for
$\cos\thetaB$, and a sum of three Gaussians with a common mean for
$\Deltaz$; the coefficients $a_0$, $a_2$, $b_0$ are close to unity
while the coefficient $b_2$ is close to zero. The $\Delta z$
likelihood distribution for continuum is determined from data in the
sideband region $5.20\GeVcc<\Mbc<5.24\GeVcc$, $|\DeltaE|<0.3\GeV$; we
treat the two subsets of the data separately, as the vertex resolution
is improved in the subset with the 4-layer SVD.  The $\calF$ and
$\cosB$ likelihood distributions are determined from MC samples.

We form product likelihoods $\calLs$ and $\calLc$ for signal and
continuum background, respectively, from the likelihood distributions
for $\calF$, $\cosB$ and (where available) $\Deltaz$, since these
quantities are independent.  In the plane $(r,\calR)$ defined by the
tagging quality $r$ and the likelihood ratio $\calR=\calLs/(\calLs +
\calLc)$, signal tends to populate the edges at $r=1$ and $\calR=1$,
while continuum tends to populate the edges at $r=0$ and $\calR=0$.
We select the events in a signal enriched region defined by $\calR >
\calR_1$ for $r>r_1$, and $\calR > 1 - \alpha(1 + r)$ for $r_2<r<r_1$,
where the parameters $r_1$, $r_2$, $\calR_1$ and $\alpha$ are mode-
and subset-dependent and are determined so that $N_S/\sqrt{N_S+N_B}$
is maximized (we use this quantity instead of $N_S/\sqrt{N_B}$ because
of the limited size of the MC simulation sample used in this
procedure).  The values are $r_1=0.85$, $r_2=0.01$, $\calR_1=0.8$ and
$\alpha=0.025$ for the first subset of the $\rhoZG$ mode; similar
values are used for the other subsets.  We define the rest of the area
as the background enriched region.


We consider the following $B$ decay backgrounds: $\BtoKG$, other $B\to
X_s\gamma$ processes, $B\to\rho\piZ$ and $\omega\piZ$, $B\to\rho\eta$
and $\omega\eta$, $B^-\to\rho^-\rho^0$, other charmless $B$ decays, and
the $b\to c$ transition.  We find the $b\to c$ background to be negligible.
The $\BtoKG$ background may mimic the signal decay $\BtoRG$ if the kaon
from $K^*$ is misidentified as a pion. To suppress
$\BtoKG$ events we calculate $\MKpi$, where the kaon mass is assigned to one
of the charged pion candidates, and reject the candidate if $\MKpi<0.96$
($0.92$) $\GeVcc$ for the $\rhoZG$ ($\rhoMG$) mode.  The decay chain
$\BtoKBG$, $\Kbar^{*0}\to\KS\piZ$, $\KS\to\piP\piM$ has a small
contribution to $\BtoOG$ due to the tail of the $K^*$ Breit-Wigner 
lineshape. In addition, $\BtoKG$ and other $\BtoXsgamma$ decays contribute
to the background when the $\rho$ and $\omega$ candidates are selected
from a random combination of particles.

Charmless decays $B\to\rho\piZ$, $\omega\piZ$, $\rho\eta$, and
$\omega\eta$ may mimic the signal if a photon from $\piZ$ or
$\eta\to\gamma\gamma$ decay is soft and undetected by the $\piZeta$
veto.  To suppress this background, we reject the candidate if
$|\coshel|>0.8$ ($0.6$) for the $\rhoZG$ and $\omegaG$ ($\rhoMG$)
modes, where the helicity angle $\thetahel$ is the angle between the
$\piP$ and $B$ momentum vectors in the $\rho$ rest frame, or the angle
between the normal to the $\omega$ decay plane and the $B$ momentum
vector in the $\omega$ rest frame.  The decay $B^-\to\rho^-\rho^0$,
$\rho^-\to\pi^-\pi^0$ also contributes to the $\BtoRZG$ mode when a
pion from the $\rho^-$ decay and a photon from the $\piZ$ decay are
both soft and undetected.  Other charmless decays have very small
contributions and are considered as an additional background component
when we extract the signal yield.


The reconstruction efficiency for each mode is defined as the fraction
of the signal yield remaining after all selection criteria, where the
signal yield is determined from an extended unbinned maximum likelihood
fit to the MC sample.
Each signal distribution is modeled as the product of a Gaussian in
$\Mbc$ and a Crystal Ball lineshape \cite{bib:cbls} in $\DeltaE$ to
reproduce the asymmetric ECL energy response for $\DeltaE$.  The
background component is modeled as the product of a linear function
for $\DeltaE$ and an ARGUS function \cite{bib:argus-function} for
$\Mbc$.  The efficiencies are listed in Table~\ref{tbl:results}.
The systematic error on the efficiency is the quadratic sum of the
following contributions, estimated using control samples: the photon
detection efficiency (2.2\%) as measured in radiative Bhabha events;
charged tracking efficiency (1.0\% per track), from partially
reconstructed $D^{*+}\to D^0\piP$, $D^0\to\KS\piP\piM$,
$\KS\to\piP(\piM)$; charged pion identification (1.0\% per pion) from
$D^{*+}\to D^0\piP$, $D^0\to\KM\piP$; neutral pion detection
(4.6--7.3\%) from $\eta$ decays to $\gamma\gamma$, $\piP\piM\piZ$ and
$3\piZ$; $\calR$-$r$ and $\piZeta$ veto requirements (5.4\%) from
$B^-\to D^0\pi^-$, $D^0\to\KM\piP$; and uncertainty due to limited 
MC statistics (0.9--1.5\%).


We perform an unbinned maximum likelihood fit to the data in the
($\Mbc$, $\DeltaE$) region bounded by $|\DeltaE|<0.3\GeV$ and
$\Mbc>5.2\GeVcc$, simultaneously for the three signal modes
and the two $\BtoKG$ modes; the latter are used to determine the residual
$K^*\gamma$ backgrounds in the signal modes.  We fit the two data
subsets simultaneously, so that in total ten distributions are
included in the fit.  We define the combined branching fraction
$\Br(\BtoROG) \equiv \Br(\BtoRMG)$, assuming the isospin relation
\cite{bib:ali-1994} $\Br(\BtoRMG) = 2\tauBratio\Br(\BtoRBG) =
2\tauBratio\Br(\BtoOG)$; we use $\tauBratio = 1.086\pm0.017$
\cite{bib:pdg2004}.  We also assume $\Br(\BtoKG)\equiv\Br(\BtoKMG) =
\tauBratio\Br(\BtoKBG)$.

We describe the events in the fit region using a sum of functions for the signal,
continuum, $\KstarG$, and other background hypotheses.  Signal parameters
for $\Mbc$ and $\DeltaE$ are calibrated using $\BtoDZpi$
and $\BtoKG$ samples, respectively.
%
We use the combined branching fractions $\BtoROG$ and $\BtoKG$ as free
parameters.
The
continuum background is modeled as the product of a linear function in
$\DeltaE$ whose slope is allowed to float, and an ARGUS function in
$\Mbc$ whose parameters are fixed from a comparison between data and MC
in the background enriched region and MC in the signal enriched region.
The continuum contribution in the data is allowed to float.  The
size of the $\KstarG$ background component in each $\ROG$ channel is
constrained using the fit to the $\KstarG$ events.  The levels of
the other backgrounds are fixed using known branching fractions or upper
limits.
%

\begin{figure}[ht]
\begin{center}
\myeps[\figtwoscale]{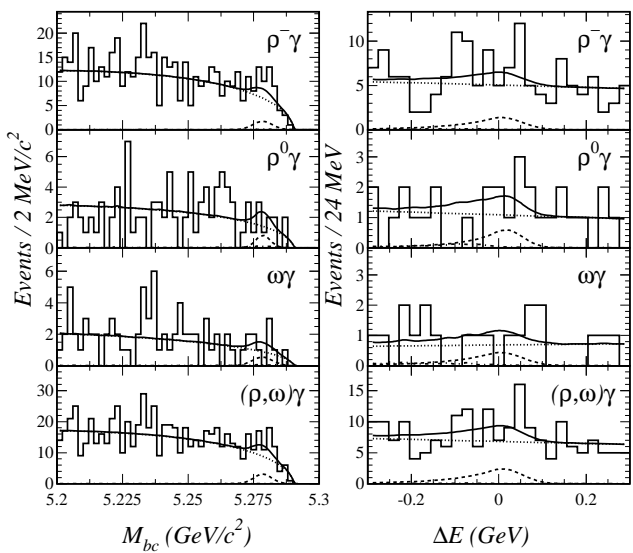}
\caption{Projections of the simultaneous fit results to $\Mbc$ (in the
         region $-0.10\GeV<\DeltaE<0.08\GeV$) and $\DeltaE$ (in the
         region $5.273\GeVcc<\Mbc<5.285\GeVcc$) for the individual modes
         and their sum.  Lines represent the total fit result
         (solid), signal (dashed), continuum (dotted), and $B$ decay
         background (dot-dashed) components.}
\label{fig:simfit}
\end{center}
\end{figure}

Results of the fit are shown in Fig.~\ref{fig:simfit} and listed in
Table~\ref{tbl:results}. The significance of the fit,
$\sqrt{-2\ln(\Lzero/\Lmax)}$, is found to be $\sROG\sigma$; where
$\Lmax$ ($\Lzero$) is the maximum likelihood from the fit when the
signal branching fraction is floated (set to zero).  To include the
effect of possible systematic error in this calculation, we change
each parameter by $\pm1\sigma$ in the direction that gives the smaller
resulting significance.  The systematic error in the signal yield is
estimated by varying each of the fixed parameters by $1\sigma$, and
then taking the quadratic sum of the deviations in the signal yield
from the nominal value.  The combined branching fraction is
$\Br(\BtoROG)=\BrROG$, where the first and second errors are
statistical and systematic, respectively.

Since the significance is small, we quote a 90\% confidence level
upper limit $\Br_{90}$ using the formula $\int_0^{\Br_{90}}
\calL(x)dx=0.9\int_0^\infty \calL(x)dx$, where $\calL(x)$ is the
likelihood function with the branching fraction fixed at $x$.  The
systematic error is taken into account assuming a Gaussian
distribution.  We find
%
$\Br(\BtoROG)<\ULROG$
at the 90\% confidence level.  
We also perform individual fits to the three signal modes;
the corresponding upper limits are listed in Table~\ref{tbl:results}.

A similar fit is performed using the ratio of branching
fractions $\Br(\BtoROG)/\Br(\BtoKG)$ instead of $\Br(\BtoROG)$, so that
the systematic error partially cancels.  We find
${\Br(\BtoROG)/\Br(\BtoKG)}<\ULROGoverKG$ at the 90\% confidence level.
One can use this result to constrain $\Vtd$: for example, using the
prescription given in Ref.~\cite{bib:ali-2004},
%
 ${\Br(\BtoROG)\over\Br(\BtoKG)}=
 \left| {\Vtd\over\Vts} \right|^2
 {(1-m_{(\rho,\omega)}^2/m_B^2)^3 \over (1-m_{K^*}^2/m_B^2)^3}
 \zeta^2
 [1 + \Delta R]$
%
where the form factor ratio $\zeta=0.85\pm0.10$ and $SU(3)$-breaking
effect $\Delta R=0.1\pm0.1$, we obtain $|\Vtd/\Vts|<\ULVtdoVts$ at the
90\% confidence level.  This limit is slightly smaller than that
resulting from searches for $B_s$ mixing~\cite{bib:pdg2004}.



In conclusion, we search for the $\btodgamma$ process using a
simultaneous fit to the $\BtoROG$ and $\BtoKG$ modes.  The
upper limit we obtain is within the range of SM
predictions~\cite{bib:ali-parkhomenko,bib:bosch-buchalla} and can be
used to constrain $|\Vtd/\Vts|$.



We thank the KEKB group for the excellent operation of the
accelerator, the KEK cryogenics group for the efficient operation of
the solenoid, and the KEK computer group and the NII for valuable
computing and Super-SINET network support.  We acknowledge support
from MEXT and JSPS (Japan); ARC and DEST (Australia); NSFC (contract
No.~10175071, China); DST (India); the BK21 program of MOEHRD and the
CHEP SRC program of KOSEF (Korea); KBN (contract No.~2P03B 01324,
Poland); MIST (Russia); MHEST (Slovenia); SNSF (Switzerland); NSC and
MOE (Taiwan); and DOE (USA).


\begin{table*}[ht]
\caption{ Yield, significance (including systematic error), efficiency and fitted
  branching fraction (central value and 90\%
  confidence level upper limit) for each mode and for
  $\BtoROG$.}
\label{tbl:results}
\begin{ruledtabular}
\begin{tabular}{lccccc}
     &       &              &            
     & \multicolumn{2}{c}{Branching Fraction} \\
Mode & Yield & Significance & Efficiency 
     & Central Value & Upper Limit \\
\hline

$\BtoRMG$ & $\nRPG$ & $\sRPG\sigma$ & $\effRPG$ & $\BrRPG$ & $\ULRPG$ \\
$\BtoRBG$ & $\nRZG$ & $\sRZG\sigma$ & $\effRZG$ & $\BrRZG$ & $\ULRZG$ \\
$\BtoOG$  & $\nOMG$ & $\sOMG\sigma$ & $\effOMG$ & $\BrOMG$ & $\ULOMG$ \\
Combined  &---  & $\sROG\sigma$ &  ---      & $\BrROG$ & $\ULROG$ \\
\end{tabular}
\end{ruledtabular}
\end{table*}

\end{document}